\documentclass[10pt,letterpaper]{article}
\usepackage{opex3}
\input{my_def_OL.def}
\begin{document}

%%%%%%%%%%%%%%%%%% title page information %%%%%%%%%%%%%%%%%%
\title{Nonlocal PMD compensation in the transmission of non-stationary streams of polarization entangled photons}
%Nonlocal compensation of Polarization Mode Dispersion in fiber-optic systems using polarization entanglement

\author{Mark Shtaif$^1$, Cristian Antonelli$^2$ and Misha Brodsky$^3$}
\address{$^1$School of Electrical Engineering, Tel Aviv University, Tel Aviv, Israel 69978 \\ $^2$DIEI and CNISM, University of L'Aquila, 67040 L'Aquila,
Italy \\ $^3$AT\&T Labs, 200 Laurel Ave. S., Middletown, NJ 07748 USA}
\email{shtaif@eng.tau.ac.il} %% email address is required



\begin{abstract}
We study the feasibility of nonlocally compensating for polarization mode dispersion (PMD), when polarization entangled photons are distributed in fiber-optic channels.
We quantify the effectiveness of nonlocal compensation while taking into account the possibility that entanglement is generated through the use of a pulsed optical pump signal.
\end{abstract}

%\ocis{060.2330,060.5565} % REPLACE WITH CORRECT OCIS CODES FOR YOUR ARTICLE

%%%%%%%%%%%%%%%%%%%%%%% References %%%%%%%%%%%%%%%%%%%%%%%%%

\section{Introduction}
\noindent Applications of quantum information protocols often rely on the ability of distributing quantum entanglement among distant users, a task for which the installed fiber-optic infrastructure is the most natural candidate. Two types of photon-entanglement are particularly popular; the entanglement of polarization \cite{Zeilinger}, and time-bin entanglement \cite{Time_bin}. While the latter scheme has an important advantage of being more tolerant to fiber propagation effects, the former is attractive due to the ease with which photon polarizations can be manipulated by using standard optical instrumentation. The greatest disadvantage of the polarization-entanglement scheme stems from its inherent sensitivity to the presence of polarization decoherence mechanisms \cite{OFC2010}, the most important of which is polarization mode dispersion (PMD) \cite{Gordon&Kogelnik}. This phenomenon, which is generated by the random residual birefringence of optical fibers, introduces a frequency dependent polarization rotation of the propagating signals, a process whose effect on polarization-entanglement could potentially be devastating \cite{OL2010}. As the consequences of PMD can be very severe also in classical high-data-rate optical communications systems, many schemes for its compensation have been proposed \cite{PMD_Comp}. The same compensating schemes could in principle be applied independently to each fiber in an entanglement distribution system. On the other hand, quantum entanglement has an intriguing non-classical property that allows approaching the compensation problem in a somewhat unusual form. The combined effect of PMD in each fiber connecting the two receiving stations (conventionally referred to as Alice and Bob) can in principle be eliminated by introducing a compensator in only one of the two optical paths, for example in the path of the photon destined for Bob.
This concept, to which we refer as nonlocal PMD compensation, is similar to Franson's nonlocal compensation of chromatic dispersion \cite{Franson_PRL,Franson_2}, and it is  closely related to the concept of decoherence-free subspaces, demonstrated by Kwiat \cite{Kwiat_science} and Altepeter \cite{Altepeter_PRL} in optically birefringent crystals.

In this paper we consider the feasibility of nonlocal PMD compensation in fiber-based entanglement distribution systems. In particular, we concentrate on the case in which the entangled photons are generated in a nonlinear medium pumped by a pulsed optical source \cite{Grice1}. The regime of pulsed pumping is of particular importance to applications as many schemes for entanglement generation in optical fibers use pumps that consist of optical pulses \cite{Chi3_Kumar,Wang2,Chi3_Takesue}. In what follows, we explore the relation between the pump-pulse duration (or bandwidth), the bandwidth of the entangled photons and the quality of compensation.
\section{PMD as a polarization decoherence mechanism}
In its simplest manifestation, valid in the case of sufficiently narrow-band signals, the effect of PMD is to cause an incident optical pulse to be split into two orthogonally polarized components that are delayed relative to each other. The polarization states of these two components are known as the principal states of polarization (PSP) and the delay between them is called the differential group delay (DGD) \cite{Gordon&Kogelnik}. This description is commonly referred to as the first-order picture, and within its range of validity, the entire phenomenon of PMD is similar to that of pure birefringence. An important difference is that in the case of PMD both the DGD and the PSPs vary randomly with frequency and time \cite{Brodsky06}. The first order picture is valid as long as the bandwidth characterizing the spectral evolution of the PMD parameters remains much larger than the bandwidth of the signals transmitted through the fiber. The former is given approximately by the inverse of the average DGD value \cite{PMD_bandwidth}. In order to preserve the clarity of the arguments and results presented in this paper, we conduct the discussions that follow for the case in which the first-order picture applies. Nonetheless, the principles underpinning our discussions can be readily (albeit in an very cumbersome form) expanded to the most general case.

To illustrate the way in which PMD causes decoherence and then to introduce the concept of nonlocal compensation, we start by discussing the case of a perfectly stationary stream of entangled photons, generated by a continuous wave (CW) pump. Assume first that PMD is present only in Alice's arm. In this case her photon can either be detected earlier, or later than the photon registered by Bob's detector, thereby disclosing that its state of polarization has collapsed  to either the fast, or the slow PSP, respectively.
Since the receivers are insensitive to the photons' arrival times, this information is lost to the environment, resulting in decoherence and in a reduction in the degree of polarization entanglement. We may now introduce the concept of nonlocal compensation. Assume that Bob includes in his optical path a PMD element whose properties over the signal spectrum are identical to those of Alice's. His photon--being polarization entangled with Alice's--will necessarily collapse into the same PSP (slow, or fast) as Alice's photon, so that the two photons will be detected simultaneously. In this case, the times of arrival disclose no information on whether the photons passed through the slow, or the fast PSP and decoherence is avoided. The situation changes drastically when a pulsed pump is used in order to generate the entangled photons. In this case, the pump imposes an independent time reference, which enables one to assess whether the photons are aligned with the slow, or the fast PSP, based on their time of arrival relative to the pump's time reference. It is therefore apparent that in this scenario, decoherence is unavoidable and the effectiveness of the compensation process deteriorates. In the section that follows we provide a formal description of this regime and quantitatively assess the compensation efficiency.
\begin{figure}[t!]
\centering\includegraphics[width=8cm]{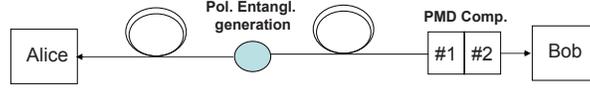}
\caption{Nonlocal PMD compensation set-up. The compensator in Bob's path contains two stages. The first compensates Bob's PMD, the second compensates for Alice's PMD.}
\end{figure}
\section{Theory and results}
Figure 1 shows a schematic description of the considered nonlocal compensation scheme. Bob's compensator is illustrated as a two-stage device, where the first stage is responsible for compensating Bob's PMD, whereas the second stage is the nonlocal compensator which compensates for the PMD present in Alice's fiber. While this description allows us to ignore the PMD that is present in Bob's arm in what follows, we point out that in a realistic implementation both stages of compensation can be realized in a single device. We consider a maximally polarization entangled two-photon state, generated by parametric down conversion in a $\chi^2$ nonlinear crystal\footnote{The same expression with small modifications to the waveform part applies to the case where entanglement is generated in a $\chi^3$ nonlinear optical medium}, which without loss of generality, can be expressed in the form\footnote{A possible phase difference between the two polarization terms, which often follows from the experimental procedure of photon generation \cite{Chi3_Kumar}, is immaterial to our analysis and is therefore omitted.}\cite{Grice1}
\bea |\psi\rip=\frac{1}{2\pi}\int\df\omega_a\int\df\omega_b H^*_a(\omega_a)H^*_b(\omega_b) \tilde E_p(\omega_a+\omega_b)|\omega_a,\omega_b\rip
\otimes \frac{1}{\sqrt{2}}\left[|\uu u_a,\uu u_b\rip+|\uu u'_a,\uu u'_b\rip\right].\label{100}\eea
The functions $H_a(\omega)$ and $H_b(\omega)$ are the transfer functions of two frequency filters positioned along Alice and Bob's optical paths, respectively, and $\tilde E_p(\omega)$ is the spectrum of the pump signal. We have assumed in Eq. (\ref{100}) that the phase matching condition is uniformly satisfied within the bandwidth of the two filters, so that the spectral dependence of phase matching is negligible. The terms $\uu u_a$, $\uu u_b$, $\uu u'_a$ and $\uu u'_b$ denote the Jones vectors describing the polarization states of Alice and Bob's photons, with the primes denoting orthogonality in polarization space, namely $\uu u_a \cdot \uu u'_a = \uu u_b \cdot \uu u'_b = 0$. The frequency dependent part in Eq. (\ref{100}) can be conveniently reexpressed as
\bea \frac{1}{2\pi}\int\df\omega_a\int\df\omega_b H^*_a(\omega_a)H^*_b(\omega_b) \tilde E_p(\omega_a+\omega_b)|\omega_a,\omega_b\rip=\int\df t_a\int\df t_b g(t_a,t_b)|t_a,t_b\rip\label{200}\eea
where $ g(t_a,t_b)=\int\df t h_a^*(t-t_a)h_b^*(t-t_b) E_p(t),$ with $h_a(t)$ and $h_b(t)$ being the impulse responses of the two filters and where $E_p(t)$ represents the complex envelope of the pump signal. As is apparent from Eq. (\ref{200}), $|g(t_a,t_b)|^2$ is proportional to the probability density that Alice and Bob's photons arrive at the detectors at the times $t_a$ and $t_b$, respectively. Normalization of the overall quantum state requires that  $g(t_a,t_b)$ satisfies $\int \df t_a\df t_b|g(t_a,t_b)|^2=1$. In the case of CW pumping, $E_p$ is a constant and $g(t_a,t_b)$ reduces to be a function of the time difference $\tau=t_a-t_b$, thereby making the entangled photon stream stationary \cite{Franson_PRL,Franson_2,Kwiat_science}. In this case the normalization condition must be modified to $\int \df \tau |g(\tau)|^2=T^{-1}$, with $T$ being the photo-detector's integration window.
In order to abbreviate the notation, we denote the polarization part of the state-vector by $|\psi_p\rip=2^{-1/2}[|\uu u_a,\uu u_b\rip+|\uu u'_a,\uu u'_b\rip]$, and by $|g(t_a,t_b)\rip$ the waveform related part given by Eq. (\ref{200}). This way, the representation of the overall state simplifies to $|\psi\rip = |g(t_a,t_b)\rip \otimes |\psi_p\rip$.

We now consider the case in which Bob introduces a second PMD element into his optical path in order to compensate for PMD that is suffered by Alice's photon (stage \#2 in Fig. 1). In order to conveniently describe the effect of PMD on the system, we note that the two-photon polarization state $|\psi_p\rip$ maintains its form in a subspace of representation bases. In particular, if one denotes the slow PSP in Alice's arm by $\uu s_a$, one can always find a state of polarization $\uu s_b$ such that $|\psi_p\rip$ is expressed in the $\{\uu s_a, \uu s_b\}$ basis as $|\psi_p\rip=2^{-1/2}[|\uu s_a,\uu s_b\rip+|\uu s'_a,\uu s'_b\rip]$. It is now apparent that in order to perform nonlocal PMD compensation, stage \#2 of Bob's compensator must contain a PMD element, whose slow PSP is aligned with $\uu s_b$. Then the quantum state of the photons received by Alice and Bob can be conveniently expressed as
\bea |\psi_{\mathrm{out}}\rip =\frac{1}{\sqrt 2} | g(t_a-\tau_a/2,t_b-\tau_b/2)\rip\otimes |\uu s_a, \uu s_b \rip + \frac{1}{\sqrt 2}  | g(t_a+\tau_a/2,t_b+\tau_b/2)\rip\otimes|\uu s_a', \uu s_b' \rip, \label{psi_out1}\eea
where $\tau_a$ is Alice's DGD and $\tau_b$ is the DGD in stage \#2 of Bob's compensator. It is evident in Eq. (\ref{psi_out1}) how the photons' states of polarization become coupled to their times of arrival. The degree of entanglement can be assessed by calculating the concurrence \cite{Wootters} from the density matrix that corresponds to $|\psi_{\mathrm{out}}\rip$ after tracing it over the time modes, i.e. $\rho=\int \df t_a\df t_b\lip t_a,t_b|\psi_{\mathrm{out}}\rip\lip \psi_{\mathrm{out}}|t_a,t_b\rip$. Following some algebra, the concurrence can be presented in the following form
\bea C\left( \tau_a,\tau_b \right)=\left|\int\df t_a\int\df t_b g(t_a,t_b)g^*(t_a+\tau_a,t_b+\tau_b)\right|\label{Conc1},\eea
where normalization of $g(t_a,t_b)$ ensures that in the absence of PMD (when $\tau_a=\tau_b=0$), the concurrence is equal to unity. The existence of a decoherence-free subspace demonstrated in \cite{Kwiat_science},\cite{Altepeter_PRL} follows from reducing the bandwidth of the pump signal to zero and by setting the DGD values $\tau_a$ and $\tau_b$ to be equal. In this case, as we mentioned earlier, normalization of $g(\tau)$ implies that $C=1$ and the degree of entanglement is fully preserved. Note that this is a direct consequence of the fact that both photons are either delayed (if they are in the $|\uu s_a, \uu s_b \rip$ state), or advanced (if they are in the $|\uu s'_a, \uu s'_b \rip$ state). As for a continuous wave pump there is no absolute time reference that would indicate whether they are late, or early, no information on their polarization states can be extracted from their times of arrival, thereby implying the absence of decoherence. Conversely, when the the pump signal consists of pulses of finite duration, the photons' times of arrival relative to the pump disclose information on their polarizations,
even if they reach the detectors simultaneously. %(if they arrive earlier than the pump they are aligned with the fast PSP and vice versa).
In this scenario decoherence cannot be avoided.

We now quantify the above principles by considering an example consisting of Gaussian shaped frequency filters and of a pump signal whose complex envelope is a Gaussian pulse. Namely,
$\tilde E_p (\omega) \propto \exp\left(-\omega^2/4B_p^2 \right)$ and $H_{a,b}(\omega) \propto \exp\left[-(\omega\pm\Delta\Omega)^2/4B_{a,b}^2\right]$,
where we use $B_p$ to denote the root-mean-square bandwidth of the optical pump $|\tilde E(\omega)|^2$ and the term $B_{a,b}$ denotes the root-mean-square bandwidths of Alice's, or  Bob's filters $|H_{a,b}(\omega)|^2$. The frequency axis was chosen such that the pump frequency is equal to 0 and the central frequencies of Alice and Bob's filters are symmetrically offset from it by $\pm\Delta\Omega$. In this condition Eq. (\ref{Conc1}) reduces to
\be C(\tau_a,\tau_b) = \exp\left[ -\frac 1 2 \frac{(\tau_a-\tau_b)^2B_a^2B_b^2}{B_p^2 + B_a^2 + B_b^2} \right]\exp\left[ -\frac 1 2 \frac{B_p^2(B_a^2 \tau_a^2 + B_b^2 \tau_b^2) }{B_p^2 + B_a^2 + B_b^2} \right]
.\label{Conc2}\ee
The first term on the right-hand-side of (\ref{Conc2}) accounts for the loss of entanglement due to a difference in the DGD values, whereas the second term represents the main effect of the pump bandwidth. Equation (\ref{Conc2}) reveals the relevant regimes of operation. When the pump bandwidth is small relative to the bandwidths $B_a$ and $B_b$ of Alice and Bob's photons, the second term in Eq. (\ref{Conc2}) reduces to unity. Then the overall concurrence can be made equal to 1 if the DGD in Bob's compensator is equal to Alice's DGD, i.e if $\tau_b=\tau_a$. In the opposite limit, when $B_p$ is greater than $B_a$ and $B_b$, the concurrence assumes a value that depends on the magnitude of the DGD. Provided that %$\tau_{a,b}B_{a,b}$ is not significantly greater than 1
the DGD is not significantly larger than the inverse bandwidth of Alice and Bob's filters, the concurrence in this case reduces to $C=\exp\left[ -(B_a^2 \tau_a^2 + B_b^2 \tau_b^2) /2 \right]$, a value that is independent of $B_p$.
The highest concurrence that can be achieved by nonlocal PMD compensation is given by
\bea C^{\mathrm{opt}}=\exp\left[-\frac 1 2 \frac{\tau_a^2}{B_p^{-2}+B_a^{-2}}\right],\label{Copt}\eea
and it is obtained when the compensators DGD is given by $\tau_b^{\mathrm{opt}}=\tau_a/\left(1+B^2_p/B^2_a\right)$, a value that
naturally reduces to $\tau_b^{\mathrm{opt}}=\tau_a$ when the pump bandwidth is set to $B_p=0$. On the other hand, $\tau_b^{\mathrm{opt}}$ approaches 0, when $B_p\gg B_a$. Since in most situations of practical interest, $B_p\leq B_a$, effective compensation takes place as long as the DGD value to be compensated is small relative to $B_p^{-1}$, i.e. relative to the pump-pulse duration. Notice also that for Gaussian waveforms neither the optimal value of $\tau_b$ needed for compensation, nor the compensated concurrence depend on the bandwidth of Bob's optical filter. Finally, we note that the sensitivity to small errors in the setting of the compensator's DGD is very weakly dependent upon the pump bandwidth, and thus for all practical purposes, the concurrence can be approximated as
$ C\simeq C^{\mathrm{opt}}\left(1-\delta\tau^2/\tau_0^2\right)$, with $\tau_b=\tau_b^{\mathrm{opt}}+\delta\tau$ and where $\tau_0= 2B_p^{-1}$.
\begin{figure}[t!]
\centering\includegraphics[width=9cm]{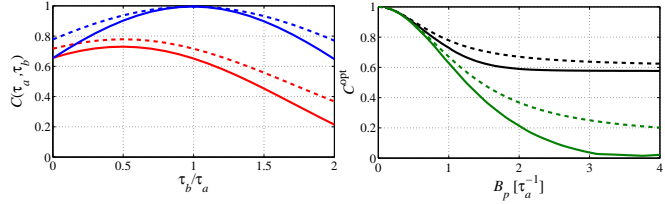}
\caption{(a) The concurrence as a function of the the DGD of Bob's compensator normalized by the DGD in Alice's optical path. The red and blue curves correspond to pump bandwidths of $B_p=0.1\tau_a^{-1}$ and $B_p=\tau_a^{-1}$, respectively. (b) The best achievable concurrence after PMD compensation, as a function of the pump bandwidth $B_p$ normalized by $\tau_a^{-1}$, where the green and black curves correspond to photon bandwidths of $B=2\tau_a^{-1}$ and $B=\tau_a^{-1}$, respectively. The solid curves are drawn for typical telecom filter-shapes (super-Gaussian), whereas the dashed curves show the analytical expressions in the Gaussian case.}
\end{figure}
Interestingly, in a short ($\sim$100Km) link, of the kind used in repeaterless quantum communications, even if the instantaneous DGD reaches values of the order of few picoseconds, compensation allows maintaining the concurrence above 0.95 for typical pump and channel bandwidths of 100GHz ($B_p = B_a = 2\pi \times$100GHz). For reference, note that acceptable quantum bit error rates in typical applications of quantum protocols usually translate into concurrence values in excess of 0.6. The above analysis assesses the maximum achievable concurrence as a function of the pump signal's bandwidth, and provides the DGD value that needs to be assigned to the compensator for optimal performance. While these results are analytical in the case of Gaussian waveforms and filters, they provide a fairly good estimate for the situation in the more realistic scenarios. To see that, we numerically evaluate the concurrence in a scenario that is consistent with the main features of a commercially available entanglement generation system \cite{Chi3_Kumar,Wang2}, which uses typical telecom filters for both the pump and the signals. As we found in measurements \cite{OFC2010}, the spectra of these filters can be accurately described by a third-order super-Gaussian profile. In all the numerical results presented in what follows, we assume that Alice's and Bob's filters have identical bandwidths $B_a=B_b=B$.

In Fig. 2a we plot the concurrence versus the DGD of Bob's compensator, for the case where $\tau_a B=1$. The results corresponding to the super-Gaussian waveforms are represented by the solid curves, whereas the dashed curves illustrate Eq. (\ref{Conc2}), for comparison. The two sets of colors correspond to two values of the pump bandwidth; the case of $B_p=0.1/\tau_a$ is shown in blue, and $B_p=1/\tau_a$ is shown in red. In the case of $B_p=0.1/\tau_a$, namely when the DGD is much smaller than the time duration of the pump pulses, perfect compensation can be achieved when $\tau_b\simeq\tau_a$. In the other case, when $B_p=1/\tau_a$, so that the pump duration is comparable to the compensated DGD, the curve peaks at a smaller value of $\tau_b$ and the optimal concurrence is visibly lower than unity. While the detailed features of the dashed and solid curves differ somewhat, the general properties are similar, as expected. Figure 2b shows the maximum achievable concurrence as a function of the pump bandwidth $B_p$, represented in units of $\tau_a^{-1}$. In this case, the two sets of curves correspond to two different bandwidths of Alice and Bob's filters; the blue curves correspond to $B=1/\tau_a$, whereas the red curves correspond to $B=2/\tau_a$. The width of the compensated concurrence functions in Fig. 2b gives an idea on the range of pump bandwidths that allows effective compensation. As can be seen from Eq. (\ref{Copt}), high quality compensation occurs when $B_p$ is smaller that the inverse DGD that is to be compensated, i.e. $B_p<\tau_a^{-1}$. Beyond that value, the remaining entanglement depends on the bandwidth of the entangled photons. The smaller the value of $B$, the higher the value to which the concurrence reduces when the bandwidth of the pump pulses is increased. This is not surprising as in the limiting case, where $B\ll\tau_a^{-1}$ the temporal uncertainty in arrival times of the entangled photons is so long, that PMD has nearly no effect on it, and the degree of entanglement remains close to unity. In this limit, however, PMD compensation is of little relevance.
\section{Conclusions}
To conclude, we have studied the concept of nonlocal PMD compensation in the context of distributing pairs of polarization entangled photons over a fiber-optic link. While perfect compensation can be achieved when the stream of photons is stationary, the quality of compensation reduces rapidly when entanglement generation schemes that rely on pulsed pumping are used. For the special case of Gaussian waveforms, we analytically found the optimal compensator setting and the resulting quality of compensation. Numerical estimation of compensation in a system with realistic telecom filter shapes produced results close to the analytical ones.
\end{document}